\documentstyle[twocolumn,aps]{revtex}
\begin{document}
\draft
\title{Melting of the classical bilayer Wigner crystal: 
 influence of the lattice symmetry}

\author{I. V. Schweigert \cite {d:gnu},
V. A. Schweigert \cite {*:gnu} and   F. M.  Peeters \cite 
{f:gnu}}
\address{\it  Departement Natuurkunde, Universiteit 
Antwerpen (UIA),\\
Universiteitsplein 1, B-2610 Antwerpen, Belgium}
\date{\today}
\maketitle

\begin{abstract}  
The  melting transition of the five different lattices of a 
bilayer crystal is 
studied using the Monte-Carlo technique. 
We found the
surprising result that the 
square lattice has a substantial larger melting 
temperature 
as compared to the other lattice structures, 
which is a consequence of the specific topology of the 
temperature induced defects. 
A new  melting criterion is formulated which we show to 
be  universal  for bilayers  as well as for single layer 
crystals.

\end{abstract} 

Wigner crystallization of electrons on the surface of liquid 
helium was first observed experimentally by Grimes 
and Adams \cite {Grimes}. In the same year, 
Nelson, Halperin 
\cite{Nelson}, and Young \cite {Young} developed a theory for a 
two stage 
continuous melting of a  two dimensional (2D) crystal which was 
based on 
the ideas of Berenzinskii \cite {Berenzinskii}, Kosterlitz and 
Thouless \cite{Kosterlitz}. 
Whether melting of a 2D crystal 
is a first order transition and proceeds discontinuously  as 
described by the theories of Kleinert \cite{Kleinert} and Chui 
\cite {Chui}, or is a 
second order transition in which  the crystal first transits 
into a  hexatic phase retaining quasi-long-range 
orientational order  and then melts into an isotropic fluid, 
is still an open question and a controversial issue.
These studies of the melting transition of a 2D systems were 
directed to  single layer crystals, 
which have  the hexagonal symmetry. This is the most 
energetically 
favored structure for potentials of the form $1/r^n$ \cite 
{Gann}. Disorder will influence Wigner crystallization as was 
demonstrated recently in Refs. \cite {Chui95}.

In recent experiments on dusty plasmas \cite {Hayashi} and 
on ion plasmas \cite {Mitchell} few layer and bilayer crystals 
were observed. Bilayer  systems exhibit a much richer crystal 
structure 
(five different lattice types) as function of the 
inter-layer distance. This allows us to study the influence 
of the lattice symmetry on melting. 
Previously, the different types 
of lattices and structural transitions in a multilayer crystal 
at $T=0$ with parabolic confinement was analysed in \cite  
{Dubin,Totsuji}.  
Different classes of lattices of the double-layer crystal were 
specified in \cite {Falko} and in \cite {Goldoni}  the stability 
of the 
classical bilayer crystal was analysed in the harmonic 
approximation.

In this letter we  study the  melting of  
a classical bilayer crystal, using   
the 
Monte Carlo (MC) simulation technique. In the crystal 
phase the particles are arranged into two parallel layers  in 
the $(x,y)$--plane which are a distance $d$ apart in the 
$z$--direction. The layers contain equal density of particles 
$n/2$ and have close packed symmetry. A single layer crystal is 
a 
limiting 
case of a bilayer crystal with $d=0$ and particle 
density $n$.
  
We  assume that the particles interact through an isotropic  
Coulomb ($\kappa =0$) or screened repulsive potential
\begin{equation}
\label{eq1}
V({\vec {r}}_i, {\vec {r}}_j) =\frac {q^2}{\epsilon 
\mid {\vec {r}}_i-{\vec {r}}_j\mid}
\exp (-\kappa {\mid {\vec {r}}_i -{\vec {r}}_j\mid}),
\end{equation} 
where $q$ is the particle charge,  $\epsilon $ the dielectric 
constant,
  $\vec {r}=(x,y,$($z$=$0,d))$
 the position of the particle, 
and $1/\kappa $ is the screening length. 
 The type of lattice symmetry  at $T=0$ depends on the 
dimensionless parameter $\nu =d/a_0$, where $d$ is the 
interlayer 
distance and  $a_0 =1/\sqrt {\pi n/2}$  is the mean 
interparticle distance. For the classical Coulomb system 
($\kappa =0$) there are two 
dimensionless parameters $\nu $  and $\Gamma =q^2/a_0 
k_BT$ which determine  the state of the system. The 
classical Yukawa system ($\kappa \neq 0$) at $T\neq 0$ is 
characterised by three independent dimensionless 
parameters: $\nu $, $\Gamma $ and $\lambda =\kappa 
a_0$. Below we measure the temperature in units of 
$T_0=q^2/a_0k_B$ and the energy in $E_0=k_BT_0$. 

The initial symmetry of the 
structure is set by the primitive vectors, the values of 
which  are derived from a calculation of the  minimal 
energy configuration for fixed $\nu $. In \cite {Goldoni} it was 
found  that the bilayer Coulomb 
crystal exhibits  five different types of lattices as  
function of the interlayer distance at $T=0$: 
 $\nu <0.006$-- hexagonal, $0.006<\nu 
<0.262$--rectangular, $0.262<\nu <0.621$--
square,
$0.621<\nu <0.732$--rhombic,  and $\nu 
>0.732$--hexagonal. Using  the standard  Metropolis 
algorithm \cite {Metropolis}  
we allow the system to approach its  
equilibrium state at  some temperature $T$, after 
executing $10^4\div 5\times 10^5$ `MC steps'. Each MC step is formed by a random 
displacement of one particle. If the new configuration has a smaller energy it 
is accepted and if the new energy is larger the configuration is accepted with 
probability $\delta <\exp (-\Delta E/T)$, where $\delta $ is a random number 
between $0$ and $1$ and $\Delta E$ is the increment in the energy.
In our numerical calculations the number of particles $N$ may change for 
different types of  bilayer crystals, but the particle density remains the same. 
We took   fragments of  288 to 780 particles, where the shape of the specimen 
was determined by the $T=0$ crystal structure, and used  periodic boundary 
conditions. Applying the Ewald technique 
the potential energy  is found by summation over all 
particles and their periodical images.
 
The potential energy of the system as function of temperature is 
shown in Fig.~1.
In the crystalline state the  potential energy of the system  
rises linearly with temperature and then at some critical 
temperature 
it increases very steeply. This  denotes the beginning of 
melting   
and is 
related to 
the  unbinding of dislocation pairs, which we will discuss 
below.  
The square bilayer crystal   
($\nu =0.4$) exhibits a jump in the potential energy at melting 
of 
size 
$\delta _e=0.71\times 10 ^{-2}k_BT_0$, and which is about a 
factor 2 
larger than for a hexagonal lattice, i.e.  at $\nu =0$,  
$\delta _e=0.39\times 10 ^{-2}k_BT_0$,   and at $\nu =0.8$,  
$\delta _e=0.31\times 10 ^{-2}k_BT_0$.   Moreover,  the square 
lattice 
has 
a substantial higher melting temperature,
and consequently is more stable against thermal 
fluctuations than the hexagonal lattice. 

To characterize the order in the system 
we calculate the  bond--angular order factor in each   layer 
\cite {Halperin} 
\begin {equation}
\label{eq3}               
G_\theta ^i = \langle 
\frac{2}{N}
            {\sum\limits_{j=1}^{N/2}}
\frac{1}{N_{nb}}{\sum\limits_{n=1}^{N_{nb}}}
\exp (iN_{nb}\theta _{j,n}) \rangle,
\end {equation}
and the translational order factor
\begin {equation}
\label{eq4}
G_{tr}^i= \langle
\frac{2}{N}
{\sum\limits_{j=1}^{N/2}}
\exp (i\vec G\cdot ({\vec {r}}_i-{\vec {r}}_j)) \rangle ,
\end {equation}
where index $i=1,2$ refers to the top and the bottom layers, 
respectively, and the total bond-angular order factor of the 
bilayer crystal is defined as 
 $G_\theta =(G_\theta ^1+G_\theta ^2)/2$ and similar for  
$G_{tr}$. 
$N_{nb}$ is the number of nearest neighbour 
particles ($N_{nb}=6,4$ for the hexagonal  and square 
lattices, respectively), $\theta _{j,n}$ is the angle between 
some 
fixed axis and the vector which connects the $j$th particle 
and its nearest $n$th neighbour, and $\vec G$ is 
a reciprocal--lattice vector. 

From the behaviour of the order factors we can  
derive the temperature at which order is lost in the system. As 
seen from Fig.~2(a) the translational and
orientational order is lost at the same temperature.  
Our numerical results show that for all five types of lattices 
the 
bond-angular order factor: 
1) decreases linearly with increasing temperature (except very 
close to the melting temperature), and 
2) it drops to zero just after it 
reaches the value 0.45. 
We found that $G_\theta $ exhibits a universal behavior as shown 
in Fig.~2(b). We checked this   
for the bilayer crystal with screened and unscreened 
Coulomb interaction and for 
a single layer crystal with a Lennard-Jones $V= 
1/r^{12}-1/r^6$   and a repulsive $V = 1/r^{12}$ 
interaction potential. 
From the present numerical results for $G_{\theta}$ we formulate  
a new criterion for melting which we believe
is universal: {\it melting occurs when the bond-angle 
correlation 
factor becomes $G_{\theta} \approx 0.45$}.  

Given the above mentioned criterion for melting we calculated 
the 
melting 
temperature using the harmonic approximation.
Therefore, we numerically diagonalized
the Hessian matrix \cite {Schweig95} for the finite
fragment of the ideal structure at zero temperature of the 
crystal with periodical boundary 
conditions in order to obtain the eigenvalues. We checked that
an increase of the size of the crystal fragment does not change 
our results.
The melting temperature is then derived by linear 
extrapolating $G_{\theta }$ to the value $0.45$.  In this way we 
obtained analytically $T_{mel}$ for different types of lattices  
which agrees with our MC 
calculations within $10\%$.

Our results for the melting temperature are summarised in the 
phase diagram of Fig.~3 where we show the melting 
temperature as a function of  $\nu $ for two different values of 
the screening parameter: $\lambda =0$ for a Coulomb inter-
particle 
interaction and $\lambda =1$ for a screened Coulomb interaction. 
For $\nu =0$ and $\lambda =0$ we obtained the well-known value 
for the 
critical $\Gamma  =132$, resulting in $T_{mel}=0.0076T_0$. 
This critical value was first measured in Ref.\cite {Grimes} and 
found to be 
$137\pm15$. 

As seen in Fig.~3 the hexagonal (I and V), rectangular (II) and 
rhombic (IV)
lattices melt at  almost the same temperatures.
Further increasing the inter-layer distance we found that for 
$\nu \simeq 3$ we obtained $T_{mel}\approx T_{mel}(\nu =0)/\sqrt 
2$. 
For the square bilayer crystal (phase III)  the melting 
temperature 
increases up to $T_{mel}=0.01078T_0$ with rising $\nu $ and 
only for $\nu >0.4$ we found that 
$T_{mel}$ starts to decrease with increasing $\nu $.
It is surprising that the square lattice has a substantial 
larger 
melting temperature
than the other lattices. This is true for Coulomb ($\lambda = 
0$)
inter-particle interaction as well as for screened Coulomb. 

The  detail analysis of the melting of the crystal 
 in the vicinity of the structural phase boundary is 
much more  complicated due to the softening of a phonon 
mode as shown in Ref. \cite {Goldoni} and is left for future 
work.
To understand why the square lattice bilayer crystal  has a 
considerable larger melting temperature, we investigated 
various temperature induced isomers of a single layer crystal 
and
compared them with those of the square lattice bilayer crystal 
with 
$\nu 
=0.4$ which has the largest  melting temperature. 
For bilayer crystals the topology 
of the defects is viewed as being composed by the top and 
the bottom staggered layers. Note, that  the energy of the 
defects which occurs in the square lattice depends on the 
interlayer distance. 
At given temperature we found that during the MC simulation 
the system transits from one 
metastable state to another.  They differ by the appearance 
of isomers 
in the crystal structure which appear with 
different probabilities. We found these isomers by freezing 
instant particle configurations during our MC steps. The 
topology of the defects, their 
energy and the bond-angular and the translational order factors 
of 
these configurations are determined. 
Each point in Fig.~4(a,b) represents one configuration 
containing 
an isomer in a single layer and the square lattice bilayer 
crystals, respectively.
The qualitative behaviour of both crystals during melting  is 
similar although the 
energy of the defects in both lattice structures is 
substantially 
different. For the single layer ($\nu =0$, Fig. 4(a)), all 
isomers at $T_1=0.00756T_0$, just before melting, and at 
$T_2=0.0076T_0$ just after melting,
 were obtained. Note, that for the square lattice 
($\nu =0.4$, Fig. 4(b)),  we took 
$T_1=0.01076T_0$ and  $T_2=0.01078T_0$. 
Typical calculated defect structures obtained from  instant 
particle configurations freezed to $T=0$ are shown in 
Fig.~5(a,b) for the 
hexagonal layer and in Fig.~5(c-f) for the square bilayer 
crystal. 
First, at $T=T_1$ the quartet of bound 
disclinations (Fig.~5(a)), point defects  (Fig.~5(c,d)) and 
correlated dislocation (Fig.~5(e)) are formed. The point defects  
appear in pairs in our MC 
calculations, which are a consequence of the periodic boundary 
condition. Note that in a single layer crystal  the total energy
of a non bounded pair of a `centred vacancy' and a `centred 
interstitial' is  
$E=0.29k_BT_0$. 
In the square bilayer crystal the point defects like
`vacancy' and the `interstitial', depicted in Fig.~5(c,d), 
appear 
also in pairs and the energy of this 
unbounded pair is $E=0.315 k_BT_0$.
The disclinations bound into a quartet   and point defects 
produce only a negligible effect on the periodic lattice 
structure 
and $G_\theta =0.8\div 0.9$ and 
$G_{tr}=0.85\div 0.95$ 
(group A in Fig.~4(a,b)).
It should be noted that in spite of prolonged annealing of 
the system during $5\times 10^5$ MC steps at a 
temperature $T_1$, which is just below melting,
we did not find more complex isomers than point defects and 
quartets of disclinations.

At the temperature $T=T_2$ uncorrelated 
extended dislocations with non-zero Burgers vector and  
unbounded disclination pairs are formed 
which causes a substantial decrease of the translational order 
(group B in 
Fig.~4(a,b) and defects shown in Fig.~5(b,f)).  
At this temperature 
single  disclinations appear and  the system looses 
order, both order factors 
become small and the system transits to the isotropic fluid 
(group C in Fig.~4(a,b)). 

Fig.~4(a,b) clearly illustrates that for a square
lattice the defects which are able to destroy the translational 
and
orientational order have a substantial larger energy than those 
of a single layer crystal 
with hexagonal symmetry.   As a 
whole the localised  and extended dislocations  as well as 
disclinations in the square bilayer crystal are defects with a  
higher energy as compared to the ones in the hexagonal 
bilayer crystal. Thus, the square type bilayer crystal 
requires larger energies in order to create defects which are 
responsible for the loss of the bond-orientional and the 
translational order and thus for  melting of the crystal.

In conclusion, we studied the melting  temperature of the five 
lattice structures in a bilayer crystal and found evidence that 
the melting temperature depends on the crystal symmetry. A 
square 
lattice has a substantial larger melting temperature than e.g. a 
hexagonal lattice.  In order to understand  this result we 
investigated the defect structures responsible for melting and 
found that the defects in a square lattice have a larger energy 
as 
compared to those in a hexagonal structure and consequently 
larger 
thermal energy is required to create them. We also formulated a 
new melting criterion: in two dimensional layers and bilayers 
melting occurs when the bond-angular order factor is  $G_\theta 
=0.45$, which is independent of the functional form of the 
interparticle interaction.   

This work was supported by INTAS-96-0235 and  the 
Flemish  Science Foundation (FW0-Vl). One of us (FMP) is a 
Researcher 
Director with FW0-Vl. We acknowledge discussions with G. Goldoni 
in the initial stage of this work.

\begin {center} { FIGURES } \end {center}

FIG. 1. The potential energy as function of temperature for the 
interlayer 
distances $\nu =0$ (solid circle), 
$\nu =0.4$ (open squares); 

FIG. 2 (a) The bond-angular ($G_{\theta }$) and the 
translational ($G_{tr}$) order factors as function 
of temperature for the 
interlayer distances $\nu =0$ (circles) and $\nu =0.4$ 
(squares);
$G_\theta $: open symbols and $G_{tr}$: solid symbols. (b) The 
bond angular order factor for different interaction potentials:
 i) screened Coulomb:
 $\nu =0$ (solid squares--$\lambda =1$, open 
ones--$\lambda =3$),  and $\nu =0.4$ 
(solid circles--$\lambda =1$, open ones--$\lambda =3$), 
ii) for the Lennard-Jones potential (solid rhombics), and 
 iii) for the 
potential $1/r^{12}$ (open rhombics).

FIG. 3. The phase diagram of the bilayer Coulomb crystal 
for without screening $\lambda =0$ (open squares) and with 
screening 
$\lambda =1$ (circles). The vertical dotted lines delimit the 
different crystal structure which are depicted in the inserts
 (open symbols for the top layer and solid symbols for the 
bottom layer). The error bars denote the uncertainty 
in the temperature nearby the structural phase boundaries.

FIG. 4. The bond-angular (solid squares) and the translational 
order factors (circles) of the different defects in (a) a single 
layer crystal ($\nu =0$), and (b) in the square lattice bilayer 
system for $\nu =0.4$. 

FIG. 5 The defects in a single layer crystal: 
 (a) quartet of 
disclinations,  and (b) two unbounded disclination pairs. In the 
square lattice bilayer crystal: 
(c) `vacancy', (d) `interstitial', 
(e) correlated dislocations and (f) a pair of 
disclinations.

\newpage
\end {document}
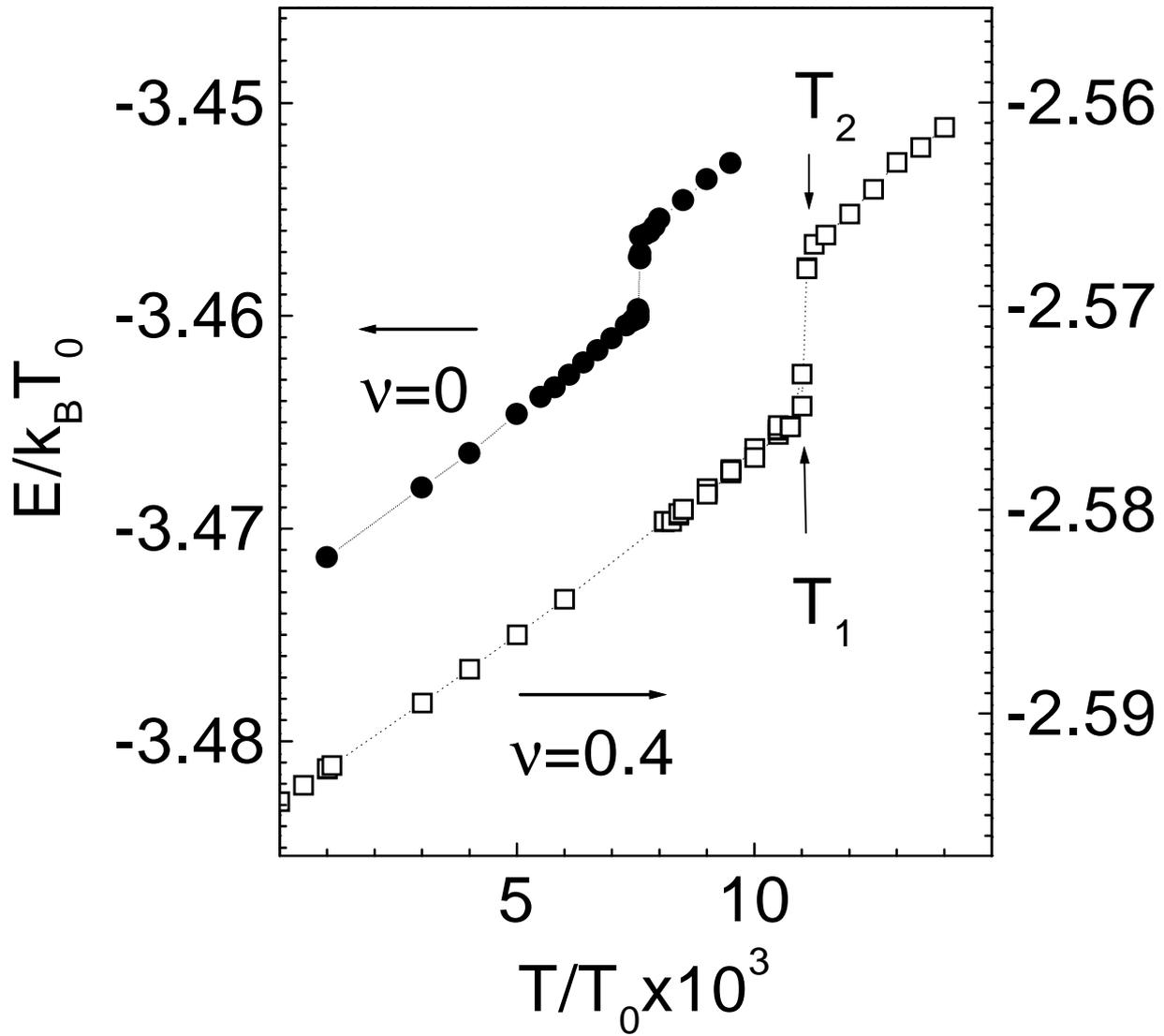
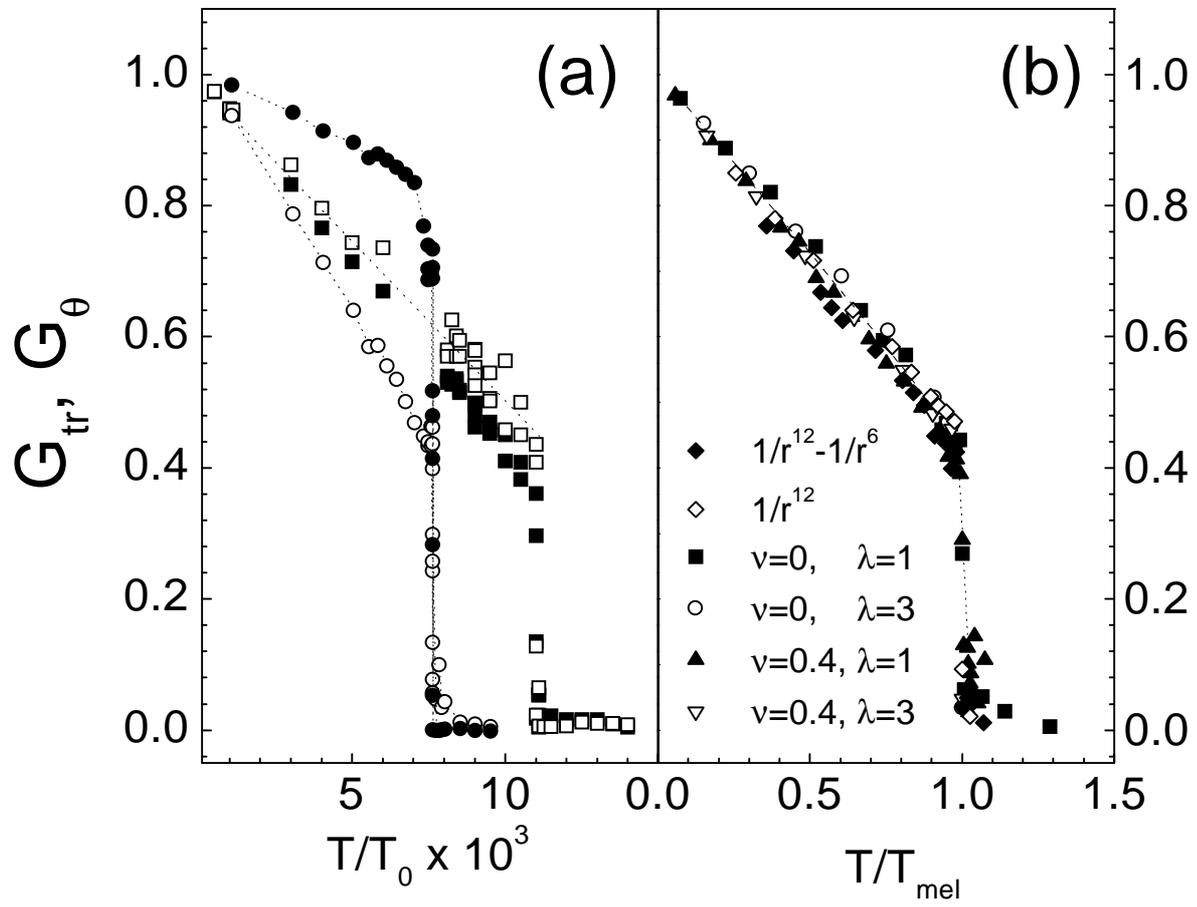
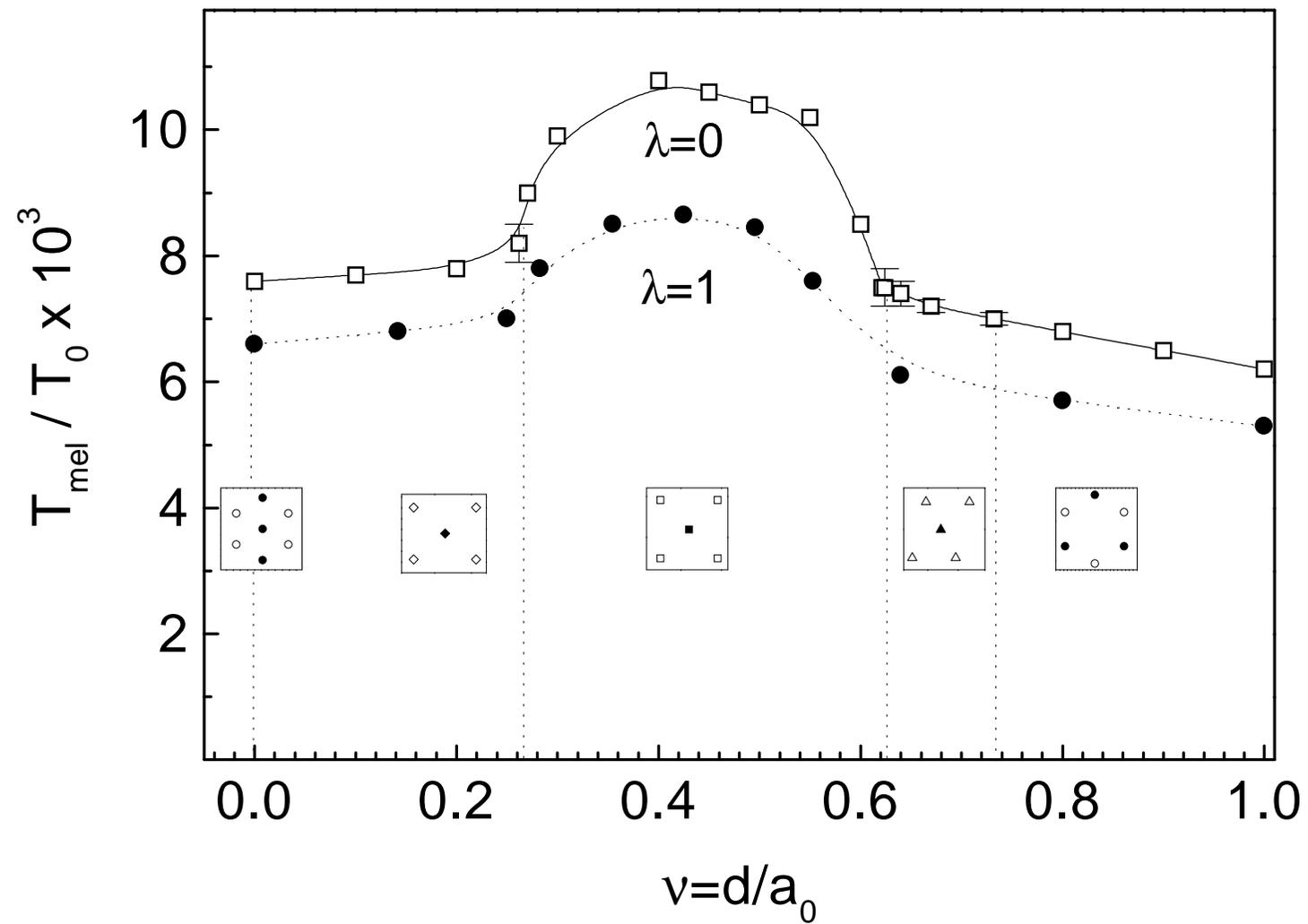
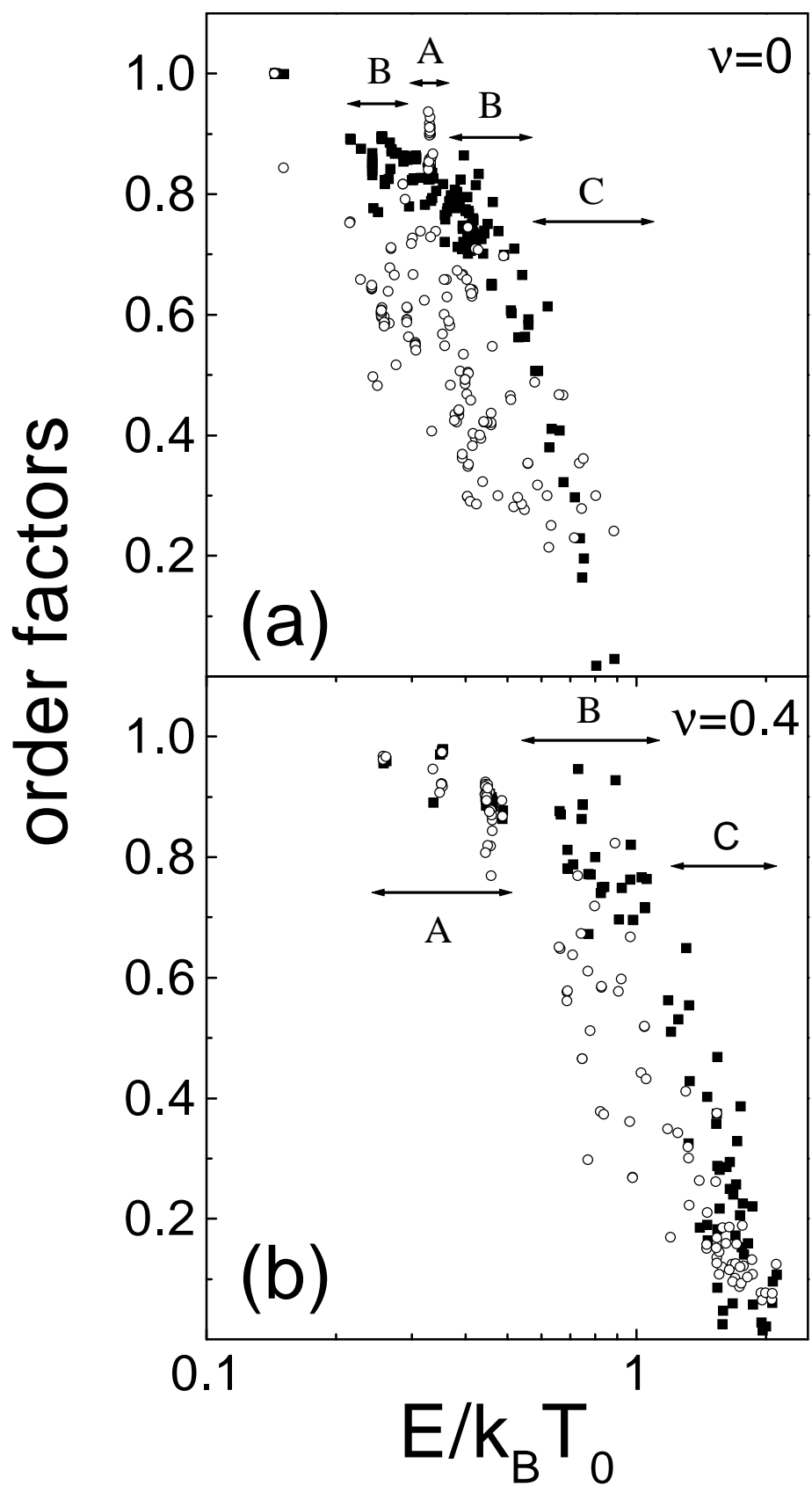
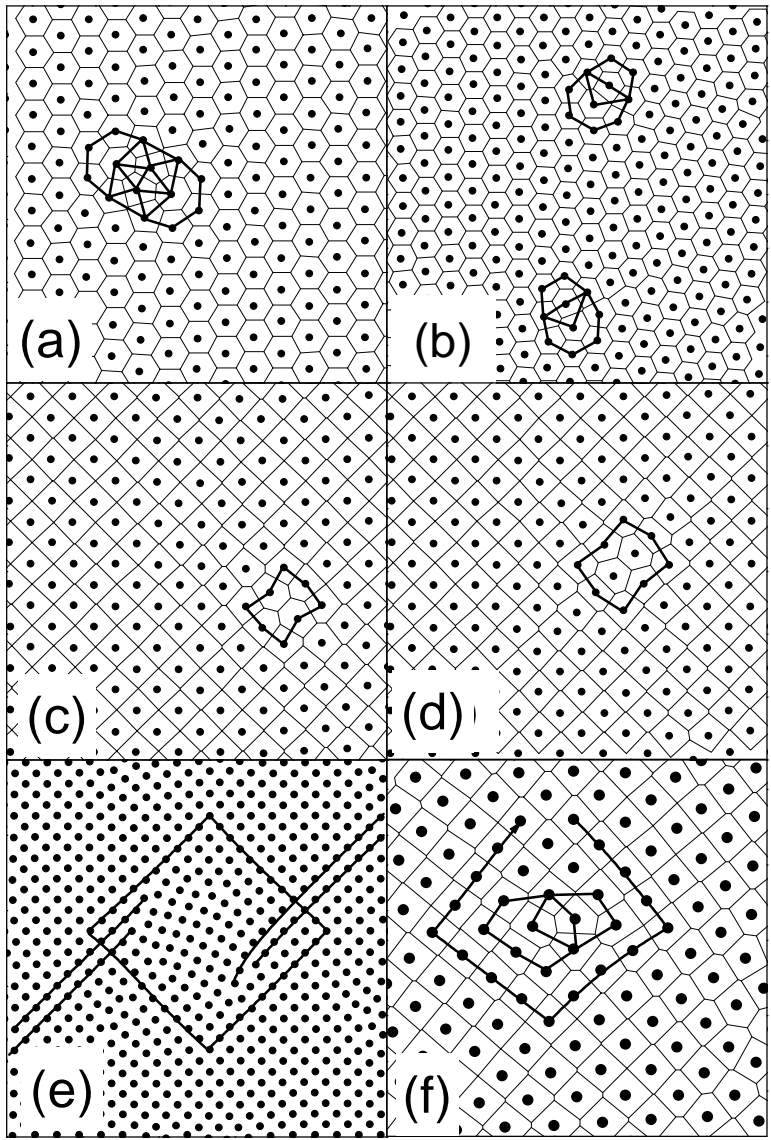
